\documentclass{PoS}
\usepackage{url}
\usepackage{bm}
\usepackage{graphicx}
\usepackage{epsfig,times,color,cite,adjustbox,wrapfig,lipsum,booktabs,comment,amssymb,amsmath}

\let\OLDthebibliography\thebibliography
\renewcommand\thebibliography[1]{
  \OLDthebibliography{#1}
  \setlength{\parskip}{0pt}
  \setlength{\itemsep}{0pt}
}

\newcommand\optin{-15pt}
\newcommand\optif{-10pt}
\title{Exploring deep learning as an event classification method for the Cherenkov Telescope Array}

\ShortTitle{Deep learning for CTA event classification}

\author{\speaker{D. Nieto}$^{1}$, A. Brill$^{2}$, B. Kim$^{2}$,
  T. B. Humensky$^{2}$, for the CTA~Consortium\\{\footnotesize
    \\$^{1}$Departamento de F\' {i}sica At\'{o}mica, Molecular y
    Nuclear, Universidad Complutense de Madrid, 28040 Madrid, Spain,
    $^{2}$ Columbia University, Department of Physics, New York, 10027
    NY, USA.}

E-mail: \email{nieto@gae.ucm.es}}

\abstract{Telescopes based on the imaging atmospheric Cherenkov
  technique (IACTs) detect images of the atmospheric showers generated
  by gamma rays and cosmic rays as they are absorbed by the
  atmosphere. The much more frequent cosmic-ray events form the main
  background when looking for gamma-ray sources, and therefore IACT
  sensitivity is significantly driven by the capability to distinguish
  between these two types of events. Supervised learning algorithms,
  like random forests and boosted decision trees, have been shown to
  effectively classify IACT events. In this contribution we present
  results from exploratory work using deep learning as an event
  classification method for the Cherenkov Telescope Array (CTA). CTA,
  conceived as an array of tens of IACTs, is an international project
  for a next-generation ground-based gamma-ray observatory, aiming to
  improve on the sensitivity of current-generation experiments by an
  order of magnitude and provide energy coverage from 20 GeV to more
  than 300 TeV.}

\FullConference{35th International Cosmic Ray Conference - ICRC2017\\
		10-20 July, 2017\\
		Bexco, Busan, Korea}

\begin{document}

\vspace{\optin}
\section{Introduction}
\vspace{\optif}

In this contribution we present an exploratory work on the application
of deep learning methods to the problem of event classification for
imaging atmospheric Cherenkov telescopes (IACTs). IACTs are capable of
imaging the particle showers created in the atmosphere by high energy
($\gtrsim 10$'s GeV) gamma rays. By focusing the Cherenkov photons
emitted by the charged particles in the shower onto a high-sensitivity
camera, images of the showers can be obtained. However, gamma-ray
initiated showers are a minor fraction of the total number of observed
atmospheric showers, which is dominated by a background of cosmic
ray-induced showers. The ability to discriminate between gamma-ray
events and cosmic-ray events is one of the major factors determining
the sensitivity of IACTs to astrophysical gamma-ray sources.

Gamma-ray initiated showers are driven by purely electromagnetic
processes, while cosmic-ray showers add hadronic processes to the
former. This difference expresses itself in the morphology of the
imaged shower and can be exploited to discriminate between the two
cases. Originally, IACT images were parametrized in terms of their
second moments~\cite{1985ICRC....3..445H}, and event classification
was performed by applying box cuts in the corresponding parameter
space. Current generation IACTs have satisfactorily implemented
classification schemes based on supervised learning algorithms trained
on a set of event-level parameters reconstructed from multiple
telescope images, as is the case of the \emph{random forest} for the
MAGIC telescope~\cite{2008NIMPA.588..424A}, and \emph{boosted decision
  trees} (BDT) for the VERITAS telescope~\cite{krause2017improved} and
the H.E.S.S. telescope~\cite{2009APh....31..383O,2011APh....34..858B},
which have substantially improved their sensitivity.

Presently, deep convolutional neural networks (CNNs), also known as
deep learning (DL), is the leading approach to supervised
representation learning, encompassing the problem of image
classification~\cite{Goodfellow-et-al-2016}. Due to its flexibility,
versatility, and performance, DL has been applied to the analysis of
data from diverse scientific disciplines, including high energy
physics (see, e.g., \cite{Baldi:2014kfa,2016JInst..11P9001A}). Thus,
it is natural to ask whether IACT image classification could benefit
from this approach, and efforts have begun to give an
answer~\cite{veritas_cnn}. One of the main advantages of DL against
previous machine learning approaches to IACT image classification is
that CNNs do not need the images to be parametrized, and therefore
have access to all the information contained in them, opening the way
to exploiting image features that might get lost or washed out during
the parametrization.

In this work, we investigate DL as an event classification method for
the future Cherenkov Telescope Array\footnote{www.cta-observatory.org}
(CTA,\cite{2013APh....43....3A}). CTA is an international project
aimed at constructing the next-generation IACT observatory, with an
order of magnitude greater sensitivity than current-generation
experiments. CTA will consist of two installations, one located in the
Northern hemisphere (La Palma, Spain) and another in the Southern
hemisphere (Cerro Paranal, Chile), each one equipped with an array of
few tens of IACTs of different sizes optimized to different energy
bands ranging from 20 GeV to more than 300 TeV. For this work we use
images from Monte Carlo simulated events as detected by an array of
9.7-m-aperture Schwarzschild-Couder medium-sized telescopes
(SC-MST,~\cite{2007APh....28...10V}) to train several known DL
architectures in the event discrimination task, leaving the
consideration of other telescope models to future works. Additionally,
we focus on single-image classification as opposed to event-level
classification (using multiple images from a same event to take
advantage of stereoscopic information), which will be dealt with in
subsequent works. Thus, this study serves as a proof of principle that
CNNs are capable of classifying air shower images, and the resulting
single-image networks will serve as a useful first stage of the more
complex array-level networks we plan to later implement.

\vspace{\optin}
\section{Deep learning}
\vspace{\optif}
\label{sec:deep_learning}

Deep convolutional neural networks (CNNs) operate fundamentally like
traditional fully connected networks, but with innovations including
local receptive fields, shared weights, and pooling that significantly
reduce the parameter count and computational cost.  A CNN consists of
many layers, each containing a set of nodes. In the input layer, each
node corresponds to one image pixel. In the following convolutional
layers, each node receives inputs from a small number of nodes in the
previous layer forming its local receptive field. These inputs are
convolved with several "filters", consisting of a grid of weights
spanning the receptive field, to produce the activations. The filter
weights are shared across each layer's nodes, a design choice that
reflects the translational and rotational invariance of many visual
features. A nonlinear activation function such as rectified linear
unit is then applied to give the node output. We refer the reader
to~\cite{deep_learning_review} for a concise overview of DL and
to~\cite{Goodfellow-et-al-2016} for a more detailed work on the
subject.

In some architectures, convolutional layers alternate with pooling
layers that combine information from neighboring nodes to reduce the
dimensionality of the feature maps and incorporate information from
wider areas of the original image ~\cite{deep_learning_review}. This
structure reflects the localized nature of image features,
dramatically reducing the needed weight parameters by ignoring
dependencies between distant pixels until later layers.

The last layer of a classifier outputs a prediction which is compared
to the true label using a loss function, such as binary cross-entropy
for binary classification. The network learns using backpropagation:
gradients of the loss surface in high-dimensional weight space are
calculated and the weights updated in the direction that minimizes the
loss using a variant of gradient descent. Training is done in batches
such that, for each epoch, the weights are updated based on the
gradient from one sample at a time, repeated until the network has
seen the entire training set.

\vspace{\optin}
\section{Application to CTA event images}
\vspace{\optif}
\label{sec:method}

We used the Monte Carlo simulation chain for CTA described
in~\cite{2013APh....43..171B}, where the atmospheric showers are
simulated with \texttt{Corsika}~\cite{1998cmcc.book.....H} and the
telescope optics and camera readout are simulated with
\texttt{Sim\_telarray}~\cite{2008APh....30..149B}. We simulated the
response of an array of 8 SCTs to $\sim5\times10^9$ proton showers and
$\sim9\times10^8$ gamma-ray showers, assuming an altitude and
atmospheric profile describing the conditions in the Roque de los
Muchachos Observatory (La Palma, Spain), where the Northern
installation of the CTA Observatory will be located\footnote{The
  Northern site of the CTA Observatory is meant to be equipped with 15
  medium-sized telescopes. The size of our array is driven by
  technical limitations of our image converter software, to be
  addressed in the future.}. The SCT model, by design, combines an
excellent point spread function across a wide field of view with a
camera featuring the highest number of pixels among CTA
telescopes. SCTs will potentially provide CTA with its
highest-resolution shower images, thus motivating the selection of SCT
images for this study. In addition, SCT gamma-ray cameras feature
square pixels, as opposed to other types of CTA telescope models whose
cameras are composed of hexagonal pixels, avoiding the need of
transforming the hexagonal lattices into 2D arrays, the common input
format for CNNs.

\begin{wrapfigure}{R}{0.5\textwidth}
  \centering
  \includegraphics[width=0.24\textwidth,clip=true,trim= 0 0 0 0]{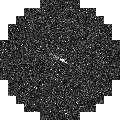}
  \includegraphics[width=0.24\textwidth,clip=true,trim= 0 0 0 0]{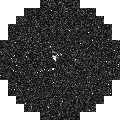}
  \caption{Some sample images from our data set, corresponding to
    independent events. \emph{Left} image from a gamma-ray initiated
    shower. \emph{Right:} image from proton initiated shower. The
    images have been normalized for better visualization.}
  \label{fig:sample_img}
\end{wrapfigure}

The energy distribution of the initial particles ranged from 3 GeV to
330 TeV in the case of gamma rays and 4 GeV to 600 TeV in the case of
protons. The arrival direction of both particle types was
homogeneously distributed inside a cone of 10$^{\circ}$ radius with
axis co-linear with the pointing position of the telescopes. Two
different pointing positions were chosen, sharing a zenith angle of
20$^{\circ}$ and Azimuth angle of 0$^{\circ}$ and 180$^{\circ}$,
splitting the simulated events evenly between the two. The output of
\texttt{Sim\_telarray} consists of the collection of digitized
photodetector pulses, for each triggered telescope camera, on an
event-by-event basis. We used the IACT analysis package \texttt{Event
  Display} to convert \texttt{Sim\_telarray} files into ROOT
format. In order to convert the raw data in ROOT format into shower
images we developed the code \texttt{ImageExtractor}, based on ROOT
and the \texttt{OpenCV} library. \texttt{ImageExtractor} performs a
simple pulse integration and stores the integrated pulse, in digital
counts, into a 120x120 array representing the camera topology that is
eventually saved into a 16-bit depth portable network graphics (PNG)
image. Relevant Monte Carlo parameters, including the particle type,
particle energy, impact parameter, and triggered telescope number, are
stored in the image header for further reference.

\begin{wraptable}{L}{0.45\textwidth}
  \centering
  \scalebox{0.9}{
    \begin{tabular}{| c |}
      \hline
      \bf{Cut} \\ \hline
      \emph{Offset}~$ \leq 3^{\circ}$ \\ \hline
      $-2 < MSCW < 2$ \\ \hline
      $-2 < MSCL < 5$ \\ \hline
      $EChi2S \ge 0$ \\ \hline
      $ERecS > 0$~TeV \\ \hline
      $0$~km~$ < $~\emph{Emission height}~$ < 50$~km \\ \hline
      $dES \geq 0$ TeV \\ \hline
  \end{tabular}}
  \caption{Arrival direction and sanity cuts applied to all data. Only
    events passing all cuts were used for training, validation, or
    testing. The cut variables are: \emph{Offset} between camera
    center and MC arrival direction of the event; \emph{MSCW (MSCL)}
    is the event's mean scaled width (length); \emph{EChi2S} is the
    $\chi^2$ its estimated vs. reconstructed energy value; \emph{ERecS
      (Emission height)} is its reconstructed energy (emission
    height); \emph{dES} is the reconstructed energy error.}
  \label{table:cuts}
\end{wraptable}

Several cuts were applied to the data before training. For all events,
the arrival direction of the shower was constrained to offsets between
$0^{\circ}$ and $3^{\circ}$ and the event telescope multiplicity was
required to be $\geq3$. Sanity cuts were also applied on several
reconstructed event parameters (see Table \ref{table:cuts}). These
cuts were chosen to match those in one of the standard CTA analysis
chains, Eventdisplay, in particular to train the default BDT-based
image classification~\cite{krause2017improved} and use the
classification performance achieved by the BDTs as a reference. The
data were separated into three energy bins corresponding to low,
medium, and high energies (see Table \ref{table:statistics}). Training
was performed separately on each energy bin.

Within each bin, the data were randomly split into training,
validation, and test sets, comprising 80\%, 10\%, and 10\% of each
bin's data. Only images in the training sets were used as inputs for
the CNN, while those in the validation sets were used to measure the
network's performance after each epoch of training. The images in the
test sets were reserved to obtain a final measure of the network's
accuracy after training was complete.  To instantiate our models, we
used the high-level neural network library
\texttt{Keras}~\cite{chollet2015keras} with \texttt{Theano}
~\cite{theano} as the computational back-end. For this work we
explored two well-known models,
\emph{ResNet50}~\cite{2015arXiv151203385H} and \emph{Inception
  V3}~\cite{2015arXiv151200567S}, both available as applications
within \texttt{Keras}.  \emph{Inception V3}, by using batch
normalization, optimization of layer dimensions, careful balancing of
network width and depth, and aggressive factorizations of the
convolutional layers, is able to achieve state of the art performance
on the ILSVRC 2012 benchmark. \emph{ResNet50}, developed in parallel
to \emph{Inception V3}, takes a different approach by using residual
mapping (implemented as "shortcut" connections across convolutional
layers) to more effectively train very deep networks. Using this
approach, networks of over 100 layers have been successfully trained
to very high accuracies.

\begin{wraptable}{R}{0.5\textwidth}
  \centering
  \scalebox{0.9}{
    \begin{tabular}{| l | l | l | l | l |}
      \hline
      Energy & $E_{min}$ & $E_{max}$ & $N_{gamma}$ & $N_{proton}$ \\ 
      bin&  [TeV] & [TeV] & &  \\ \hline
      \multicolumn{3}{|l|}{Total} & 4160578 & 4056723 \\ \hline
      Low & 0.1 & 0.31 & 727316 & 228959 \\ \hline 
      Medium & 0.31 & 1 & 657397 & 119704 \\ \hline
      High & 1 & 10 & 642034 & 72034 \\ \hline 
  \end{tabular}}
  \caption{Dataset used in this proceeding. All data were simulated at a
    Zenith angle of $20^{\circ}$ and at an Azimuth angle of either
    $0^{\circ}$ or $180^{\circ}$, split roughly evenly between the
    two. Total statistics are prior to any cut.}
  \label{table:statistics}
\end{wraptable}

We trained both architectures on our dataset and compared their
performance. The networks were not initialized with any pretrained
weights. To adapt the networks for CTA data, the input images were
resized to 240 x 240 arrays\footnote{For technical reasons, the
  networks have a minimum input width and length greater than 120, so
  the input size was doubled in each dimension.} and the output layer
replaced with a binary classifier.  Training was conducted on two
different computing systems sharing identical datasets and featuring
similarly performing GPUs: a Nvidia\texttrademark ~GeForce GTX TITAN X
Pascal and Nvidia\texttrademark~GeForce GTX 1080 Ti. The training time
over the full dataset was of 12 hours per energy bin and per model,
while the classification of data is performed approximately three
orders of magnitude faster.

\vspace{\optin}
\section{Results}
\vspace{\optif}
\label{sec:results}

\begin{figure}[t]
  \centering
  \includegraphics[width=0.45\textwidth,clip=true,trim= 0 0 0 0]{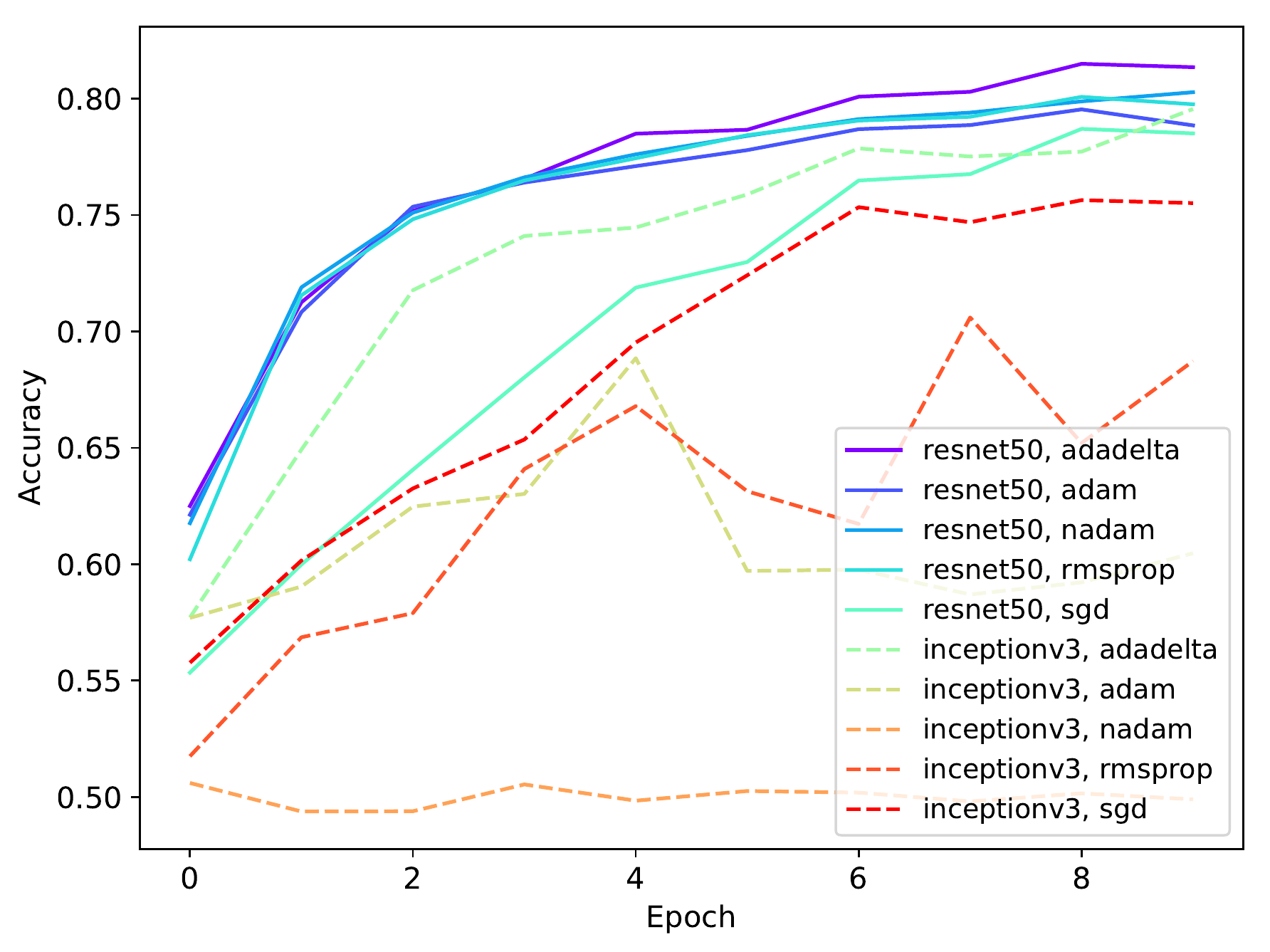}
  \includegraphics[width=0.45\textwidth,clip=true,trim= 0 0 0 0]{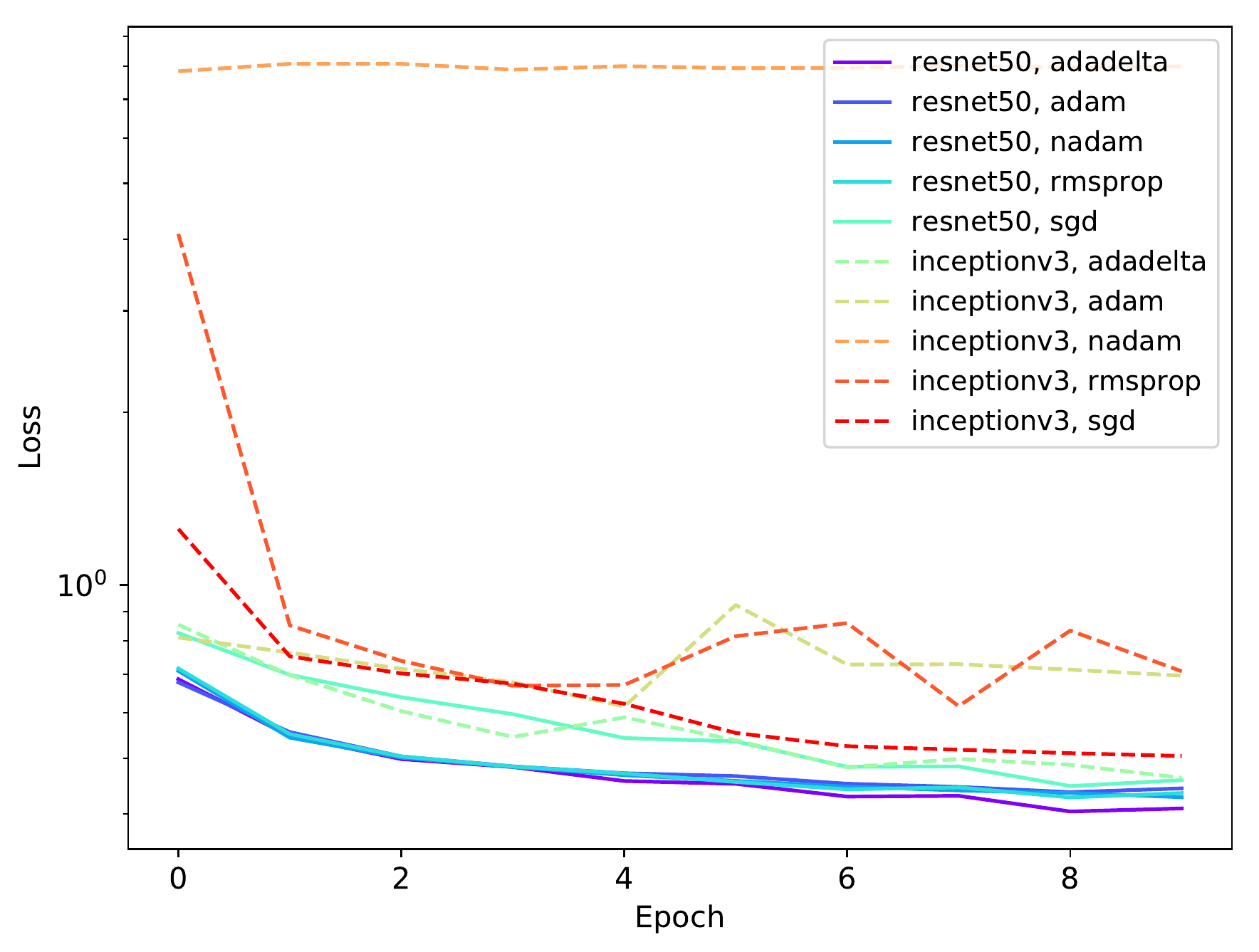}
  \caption{Network accuracy (\emph{left}) and loss (\emph{right}) as a
    function of the training epoch, for \emph{ResNet50} and
    \emph{InceptionV3} models. The training was performed on a subset
    of $2\times10^{5}$ images per category drawn from the medium
    energy bin of the training dataset.}
  \label{fig:opt_sel}
\end{figure}

The maximum accuracy achievable by training a given network depends on
the backpropagation algorithm used. We surveyed a subset of optimizers
directly available in \texttt{Keras} in order to identify the best
performing ones in terms of accuracy (fraction of images showing a
classifier score for its true category higher than 0.5, where the
classifier range is 0 to 1). Tests were conducted using stochastic
gradient descent (SGD), RMSprop, Adam, Adadelta, and Nadam
optimizers~\cite{nadam}. We trained \emph{ResNet50} and
\emph{Inception V3} on a small subset of images ($2\times10^{5}$
images per category), for a duration of 10 epochs, to evaluate the
model accuracy and loss as a function of the chosen optimizer. We did
not optimize the learning rates or any other hyperparameters of the
different optimizers, adopting the default values as provided in
\texttt{Keras}. Our findings are shown in Fig.~\ref{fig:opt_sel}. Our
results show that the default hyperparameters work reasonably well for
\emph{ResNet50} while most optimizers are not well tuned for
\emph{Inception V3}. Adadelta provided the highest accuracies at epoch
10 for both models. Consequently, we adopted Adadelta as the optimizer
for the training on the full data set, allowing for a more direct
comparison between performances.

\begin{figure}[t]
  \centering
  \includegraphics[width=0.45\textwidth,clip=true,trim= 0 0 0 0]{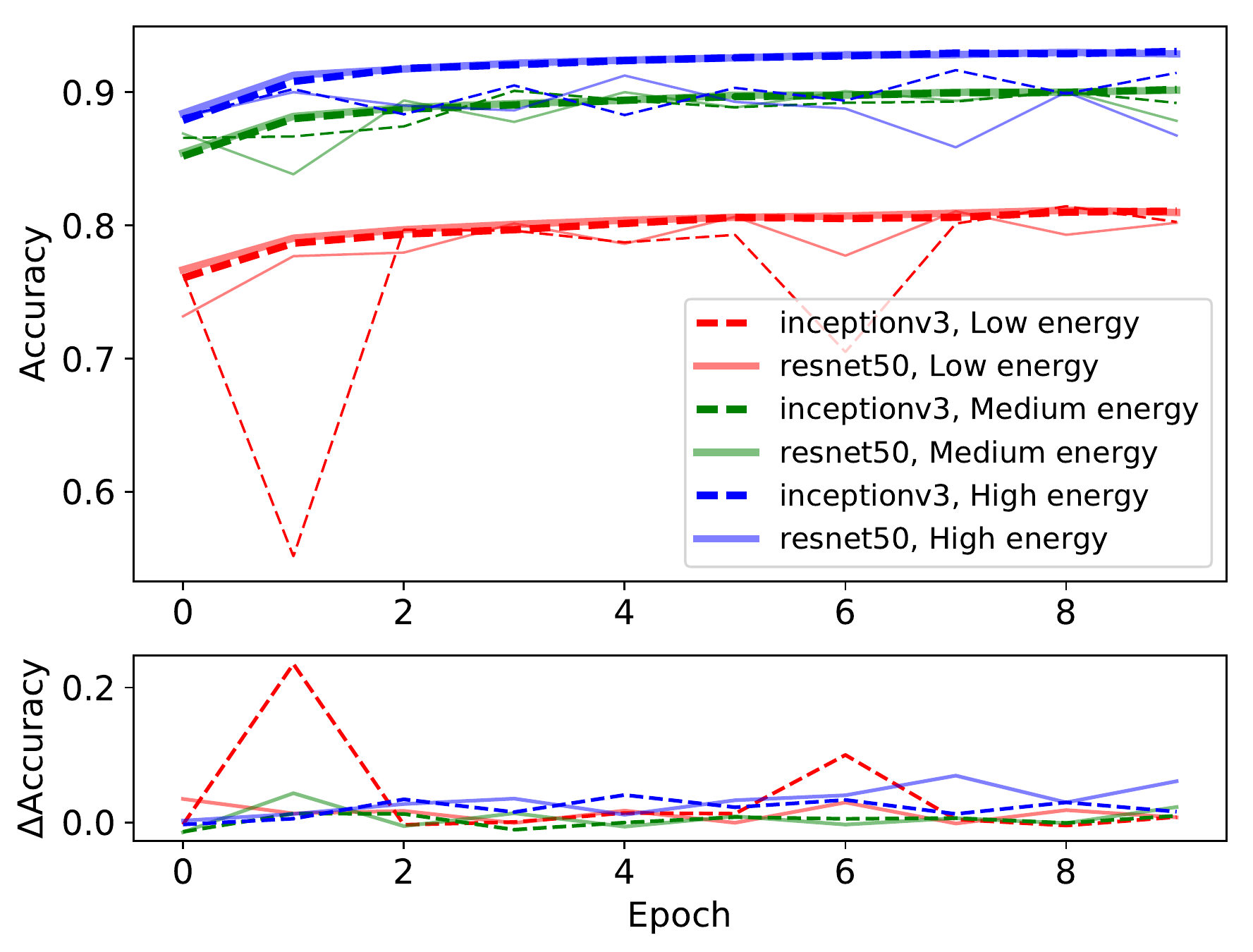}
  \includegraphics[width=0.45\textwidth,clip=true,trim= 0 0 0 0]{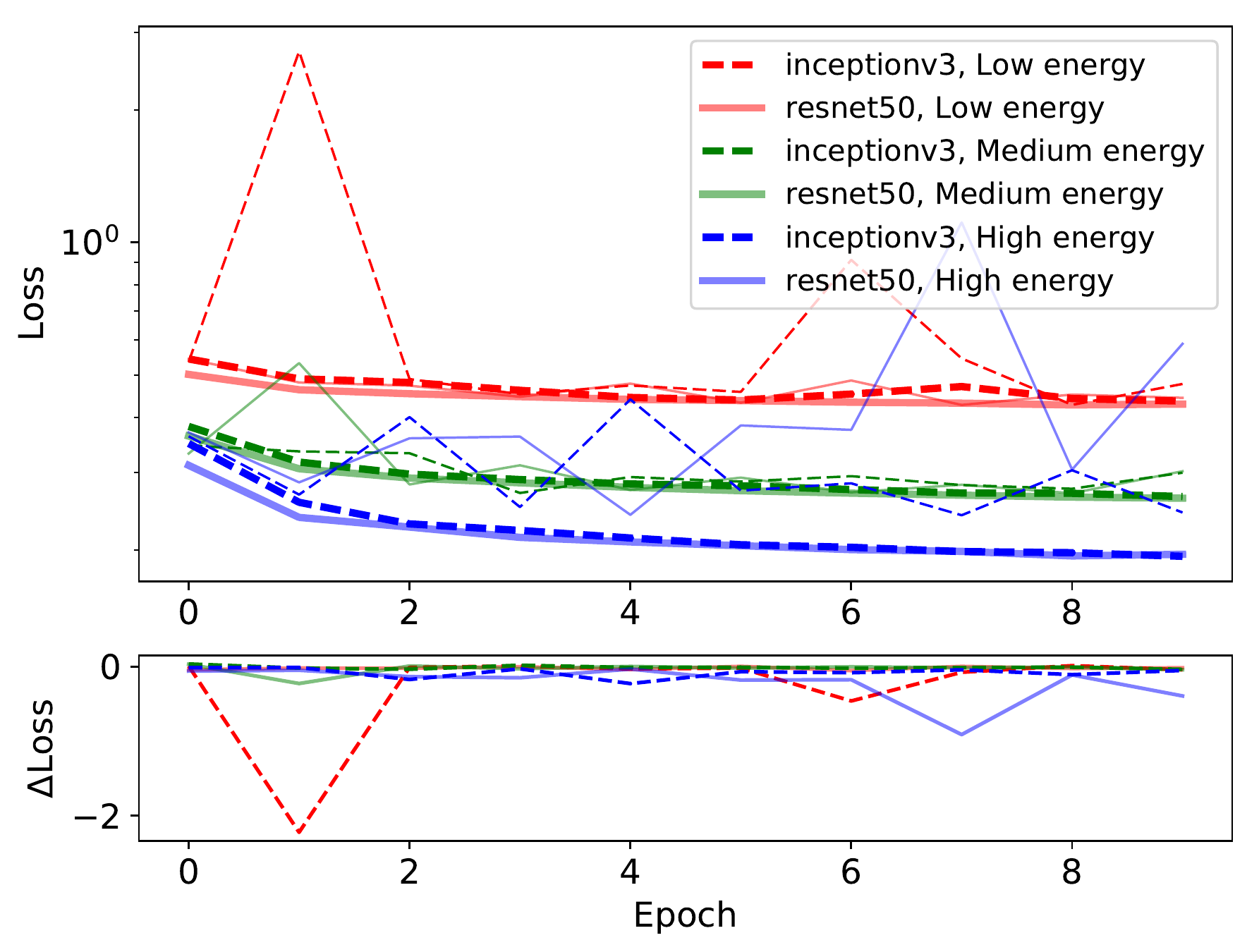}
  \caption{Network training accuracy (\emph{right}) and loss
    (\emph{left}) as a function of the training epoch, for
    \emph{ResNet50} and \emph{InceptionV3} models and all three energy
    bins. For each model and energy bin both the training and
    validation accuracy and loss are shown (upper plot), as well as
    the gap between the two (lower plot): training curves are
    represented by thick lines, while validation accuracy curves are
    represented by thin lines. The training was performed on our full
    data set, independently for each energy bin.}
  \label{fig:full_acc}
\end{figure}

After the selection of the optimizer, both \emph{ResNet50} and
\emph{Inception V3} networks were trained using the full training and
validation sets for the three energy bins. We trained our models for
10 epochs, which appeared to be enough for a stabilization of the
training accuracy without a drop in the validation accuracy that might
point out overtraining. A summary of the evolution of training and
validation accuracy as well as loss can be found in
Fig.~\ref{fig:full_acc}. We found that both models achieve a very
similar training accuracy at epoch 10, \emph{ResNet50} learning at a
slightly faster pace than \emph{Inception V3}, with the latter
providing with marginally higher validation accuracies in the low
($0.3\%$) and high ($0.4\%$) energy ranges, while tying in the medium
energy range. As expected, higher energies provide with higher
accuracies, since the images tend to be larger and brighter (more
information), thus allowing for a more effective training. A summary
of the highest validation accuracies per training can be found in
Table~\ref{table:accuracy}.

\begin{wraptable}{R}{0.5\textwidth}
  \centering
  \scalebox{0.9}{
    \begin{tabular}{| l | l | l | l |}
      \hline
      Model & Low E. & Med. E. & High E. \\ \hline
      \emph{ResNet50} & 81.1\% & 90.1\% & 91.2\% \\
      \emph{Inception V3} & 81.4\% & 90.1\% & 91.6\% \\\hline
  \end{tabular}}
  \caption{Highest classification accuracy on the validation data set
    for \emph{ResNet50} and \emph{Inception V3} models for all energy
    bins.}
  \label{table:accuracy}
\end{wraptable}

\begin{wrapfigure}{R}{0.5\textwidth}
\vspace{-1.2cm}
  \centering
    \includegraphics[width=0.5\textwidth,clip=true,trim= 0 0 0 0]{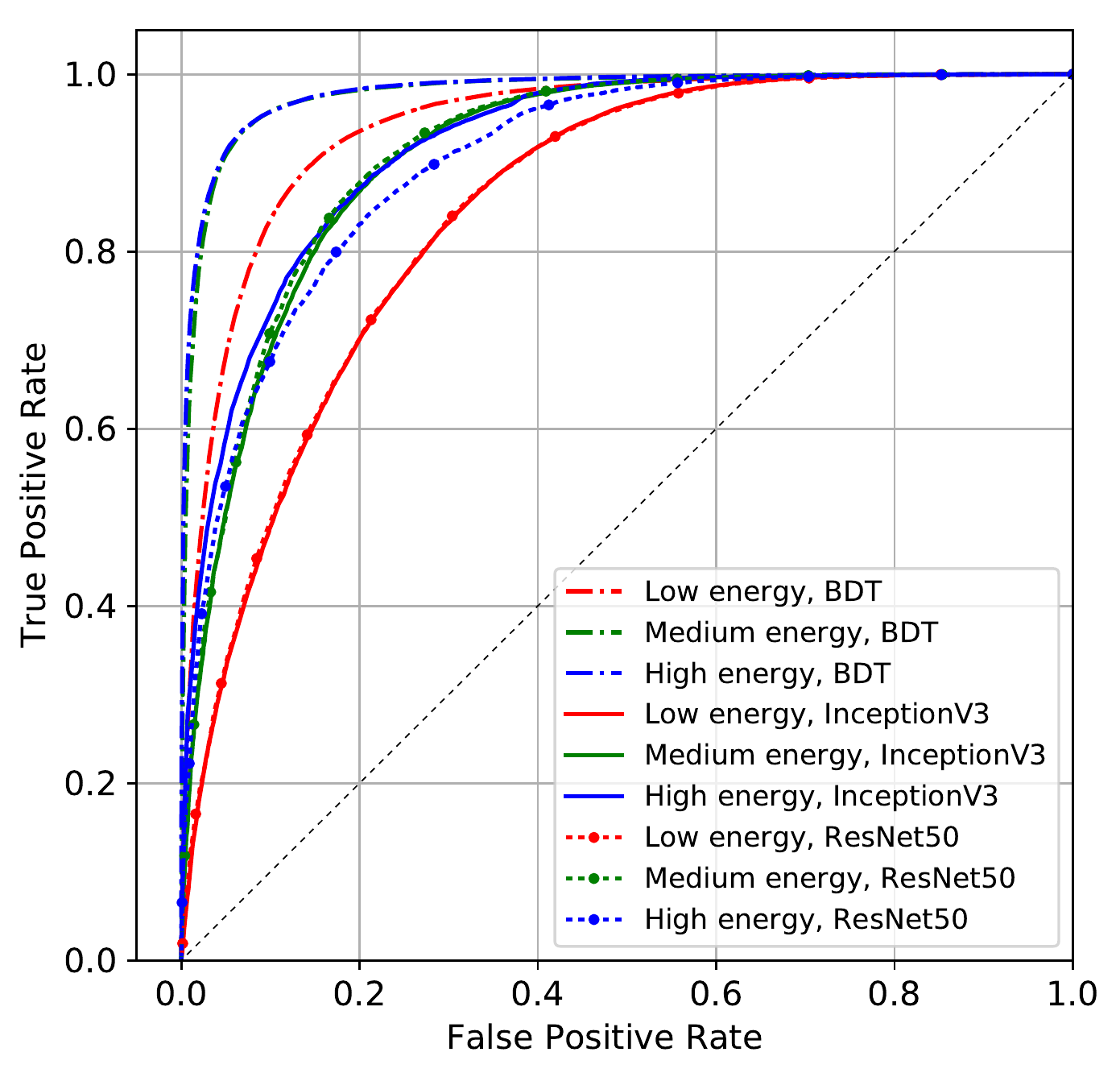}
    \caption{ROC curves from test data classified with \emph{ResNet50}
    and \emph{Inception V3} for the low, medium, and high energy
    bins. As a reference we show the ROC curves corresponding to BDT
    trained on the same data set (events parametrized using stereo
    information).}
  \label{fig:roc_comb}
\end{wrapfigure}

We generated receiver operating characteristic (ROC) curves for all
scenarios by passing the corresponding test subsets to the trained
models showing the highest validation accuracies. The collection of
all ROC curves can be found in Fig.~\ref{fig:roc_comb}. It can be seen
that, for the low and medium energies, both models are performing
almost equally. For the highest energies there seems to be some
distance between \emph{InceptionV3} and \emph{ResNet50} that we
suspect may be originated by a higher susceptibility of the former to
the relatively low image statistics present in the proton subsample,
which is hinted in the behavior of the validation and loss for the
validation sample in Fig.~\ref{fig:full_acc}. Similarly found in the
behavior of the BDTs, the \emph{InceptionV3} ROC curves for medium and
high energies are close together, although we suspect that the
discrimination power of CNNs in the highest energy range may increase
if a larger sample of proton images is used for the training of the
CNNs. While the ROC curves from the tested CNN architectures are
significantly below the corresponding ones for the BDT-based
classification method, one must bear in mind that the latter draws its
power from a multi-image event stereo reconstruction, whereas our
models are trained on individual images (coming from events that
survive the above mentioned cuts but lacking any additional stereo
information). Thus, the BDT ROC curve should be seen as a reference, a
target to overtake in the future.

\vspace{-12pt}
\section{Conclusion and outlook}
\vspace{\optif}
\label{sec:conclusion}

We have demonstrated that DL is capable of classifying simulated IACT
images, here represented by SCT images, without any prior
parametrization nor any assumption on the nature of the images
themselves. The accuracy of the tested models is energy dependent,
ranging from ~81.4\% in the low energy range to 91.6\% in the high
energy range for the \emph{Inception V3} architecture.

After applying CNNs to gamma-hadron separation on single-telescope
images, the next step is to apply these techniques to array-level
classification. IACT array analysis methods rely on combining
information from multiple telescopes to achieve high-quality stereo
shower reconstruction, perform background discrimination, and
calculate reconstructed parameters. Going forward, the focus of our
efforts will be on developing multi-input architectures consisting of
separate CNNs for each telescope, which will then combine information
through a fully-connected classifier to achieve more effective
gamma-hadron separation on the array level. Eventually, we will assess
the benefits of DL as compared to the current standard analysis on the
instrument response functions.

We also plan to further investigate different architectures, gradient
descent algorithms, and other hyper-parameters, narrowing down on the
choice that provides optimal results. It may be possible to achieve
superior results by customizing networks for the specific task of
Cherenkov image classification.  Other promising applications for CNNs
to IACT data analysis include using similar methods to those presented
here to do energy and angular reconstruction and performing more
difficult background discrimination tasks such as gamma/electron
separation or cosmic-ray composition studies.

\vspace{-12pt}
\section{Acknowledgments}
\vspace{\optif}

We gratefully acknowledge support from the agencies and organizations
listed under Funding Agencies at this website:
http://www.cta-observatory.org/. This work was conducted in the
context of the Analysis and Simulations Working Group of the CTA
Consortium. DN wants to acknowledge support from the Spanish Ministry
of Economy, Industry, and Competitiveness / ERDF UE grant
FPA2015-73913-JIN. We want to thank Tarek Hassan for discussions at
the early stages of this work. We acknowledge the support of NVIDIA
Corporation with the donation of the Titan X Pascal GPU used for this
research.

\end{document}